\documentclass[prc,preprint,showpacs,preprintnumbers,amsmath,amssymb]{revtex4}

\usepackage[mathscr]{eucal}
\usepackage{graphicx}
\def\slr#1{\setbox0=\hbox{$#1$}           
   \dimen0=\wd0                                 
   \setbox1=\hbox{/} \dimen1=\wd1               
   \ifdim\dimen0>\dimen1                        
      \rlap{\hbox to \dimen0{\hfil/\hfil}}      
      #1                                        
   \else                                        
      \rlap{\hbox to \dimen1{\hfil$#1$\hfil}}   
      /                                         
   \fi}

\def\be{\begin{eqnarray}}
\def\ee{\end{eqnarray}}

\def\etal{\emph{et al.}}

\renewcommand{\theequation}%
    {\arabic{section}.\arabic{equation}}
\makeatletter \@addtoreset{equation}{section} \makeatother

\begin{document}

\preprint{BCCNT: 05/607/335}

\title{Relativistic Model of Triquark Structure}

\author{Hu Li}
\author{C. M. Shakin}
\email[email:]{cshakin@brooklyn.cuny.edu}

\affiliation{%
Department of Physics\\
Brooklyn College of the City University of New York\\
Brooklyn, New York 11210
}%

\author{Xiangdong Li}
\email[email:]{xli@citytech.cuny.edu}
\affiliation{%
Department of Computer System Technology\\
New York City College of Technology of the City University of New
York\\
Brooklyn, New York 11201 }%

\date{\today}

\begin{abstract}

At this point it is still unclear whether pentaquarks exist. While
they have be seen in some experiments there are many experiments
in which they are not found. On the assumption that pentaquarks
exist, several authors have studied the properties of pentaquarks.
One description considered is that of pentaquarks which consist of
a diquark coupled to a triquark. There is a quite extensive
literature concerning the properties of diquarks and their
importance in the description of the nucleon has been considered
by several authors. On the other hand, there is little work
reported concerning the description of triquarks. In the present
work we study a model for the triquark in which it is composed of
a component which contains a quark coupled to a scalar diquark and
another two components in which there is a quark coupled to a
kaon. We solve for the wave function of the triquark and obtain a
mass for the triquark of 0.81 GeV which is quite close to the
value of 0.80 GeV obtained in a QCD sum rule study of triquark
properties.

\end{abstract}


\maketitle

\section{INTRODUCTION}

There has been a great deal of interest in the study of
pentaquarks and a large number of experiments have been carried
out [1-11]. The existence of pentaquarks is uncertain since they
have been seen in some experiments and not in others. Various
groups hope to clarify this situation by performing more precise
experimental searches. The $\Theta^+(1540)$ which decays to a kaon
and a nucleon has been seen in several experiments. It has been
interpreted as a pentaquark with a $udud\bar{s}$ structure [12]. A
pentaquark $\Theta_c^0$ with the assumed structure $udud\bar{c}$
has also been observed recently. A recent review may be found in
Ref. [13].

We were particularly interested in the diquark-triquark model of
Karliner and Lipkin which has been applied in the study of
pentaquarks [14,15], and we have made use of a variant of that
model in our work [16,17]. One problem for the theorist has been
the very small widths of the observed pentaquarks. We have studied
that question in a relativistic diquark-triquark model and found
we could explain the small widths seen in experiment [16,17]. In
our model, as in that of Refs. [14,15], the pentaquark is
described as scalar diquark coupled to a triquark. [See Fig.\,1.]
Using the insight gained in our analysis of the nucleon, which
made use of a quark-diquark model [18], we took the scalar diquark
mass to be 400 MeV in our study of the pentaquark.

There have been many studies of diquark structure making use of
the Nambu-Jona-Lasinio (NJL) model. Some of that work is reviewed
in Ref. [19]. One may suggest that, in addition to studies of
diquark structure, a study of triquark structure may be of
interest. In our earlier work the triquark mass was taken to be
800 MeV on phenomenological grounds. We note that a calculation of
triquark properties, using the operator product expansion and
including direct instanton contributions, obtained a triquark of
$ud\bar{s}$ structure of mass 800 MeV [20]. An additional triquark
state was found at 900 MeV in Ref. [20]. (Another work making use
of QCD sum rules yields quite small values for the width of the
$\Theta^+(1540)$ pentaquark [21].)

Once we decide to study triquark structure, we face the following
problem. The triquark of mass 800 MeV is very close to the
threshold for decay to a 400 MeV diquark and a 450 MeV strange
quark. Similarly, the triquark mass is close to the mass of a $u$
(or $d$) quark of 350 MeV and a kaon of mass 495 MeV. This
difficultly cannot be overcome by including a confinement model
since there is no confining interaction between a kaon and a
quark. Therefore, in the present work we have made use of the
quark propagator obtained in Ref. [22]. In that work we considered
quark propagation in the presence of a gluon condensate and found
that the quark propagator had no on-mass-shell poles. That is, the
quark was a non-propagating mode in the presence of the
condensate. As we will see, the use of the quark propagator of
Ref. [22] enables us to proceed in our analysis of triquark
structure. (In an early work we used the form of the propagator
discussed in Ref. [22] in a study of nontopological solitons
[23].)

The organization of our work is as follows. In Section II we
review our model of quark propagation in the presence of a gluon
condensate. In that model the quark propagator has no
on-mass-shell poles. In Section III we present the equation for
the vertex describing triquark decay to the channels: i) a $u$
quark plus a $\emph{K}^0$ meson, ii) a scalar diquark plus a
$\bar{s}$ quark and  iii) a $d$ quark and a $\emph{K}^+$ meson.
[See Figs. 2-4.] In Section IV we describe the results of our
analysis and in Section V we present some further comments and
conclusions. The Appendix contains a discussion of the
normalization of the wave functions of the scalar diquark and the
kaon.

\begin{figure}
\includegraphics[bb=0 355 519 623, angle=0, scale=0.7]{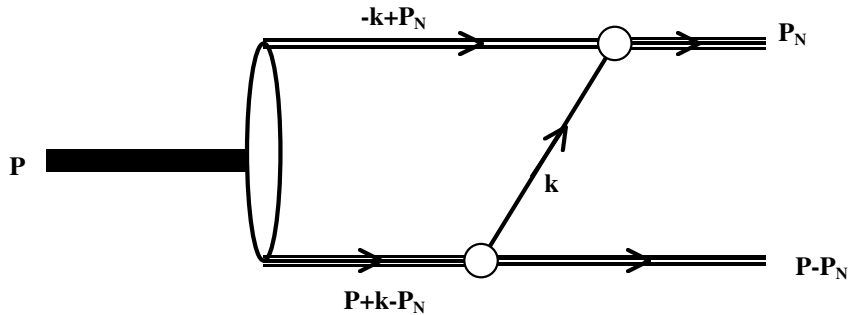}%
\caption{In this figure the heavy line denotes the pentaquark and
the line of momentum $-k+P_N$ denotes an on-mass-shell diquark.
The line of momentum $k$ represents the quark and $P+k-P_N$ is the
momentum of the triquark. In the final state we have a baryon of
momentum $P_N$ and a meson of momentum $P-P_N$.}
\end{figure}

\begin{figure}
\includegraphics[bb=20 100 765 470, angle=0, scale=0.4]{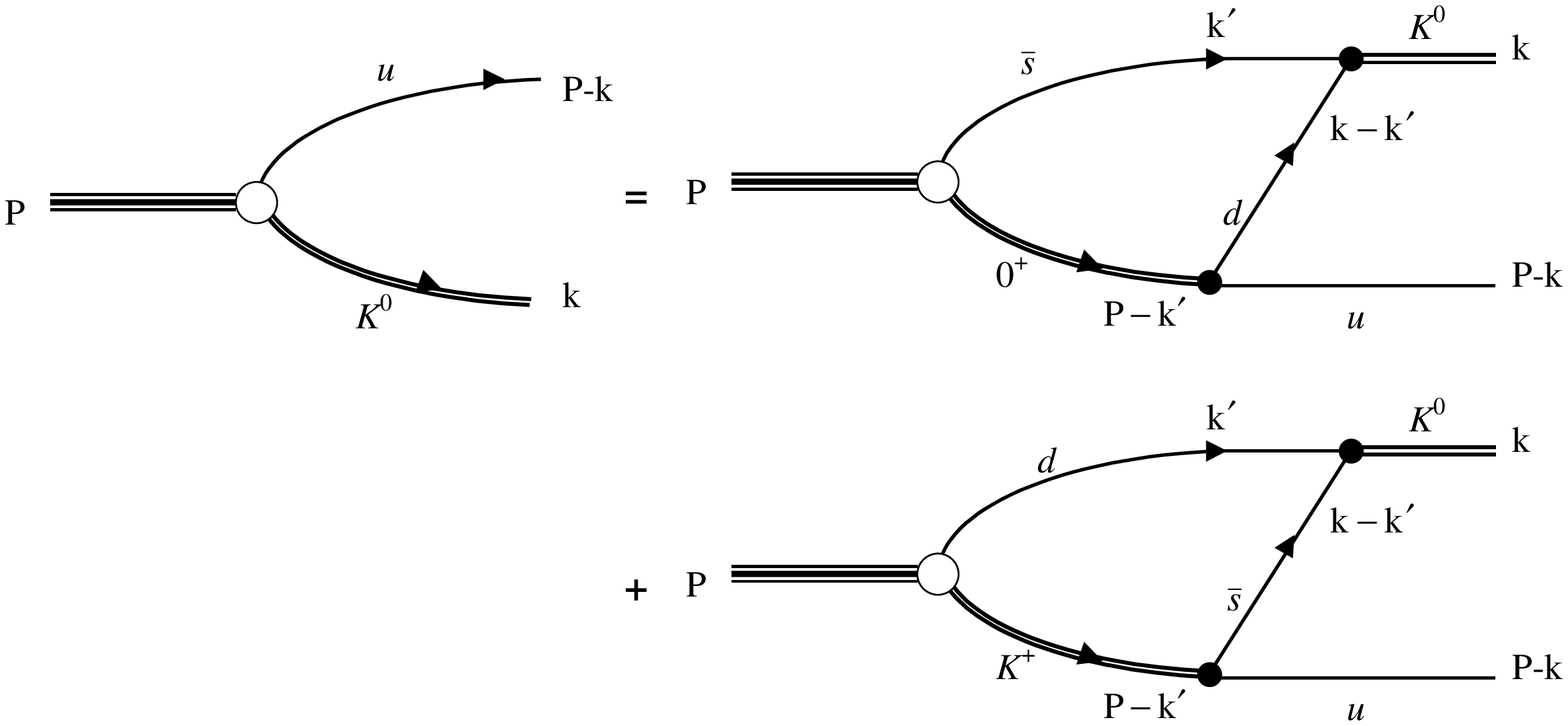}%
\caption{The equation for the triquark vertex function for the
$K^0$ and $u$ quark component.}
\end{figure}

\begin{figure}
\includegraphics[bb=20 100 765 470, angle=0, scale=0.4]{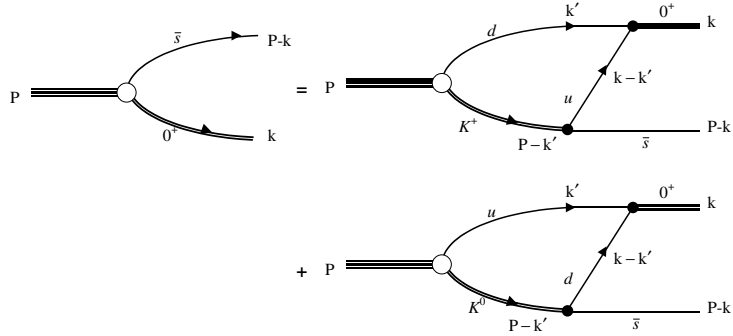}%
\caption{The equation for the triquark vertex function for the
diquark and $\bar{s}$ quark component.}
\end{figure}

\begin{figure}
\includegraphics[bb=20 100 765 470, angle=0, scale=0.4]{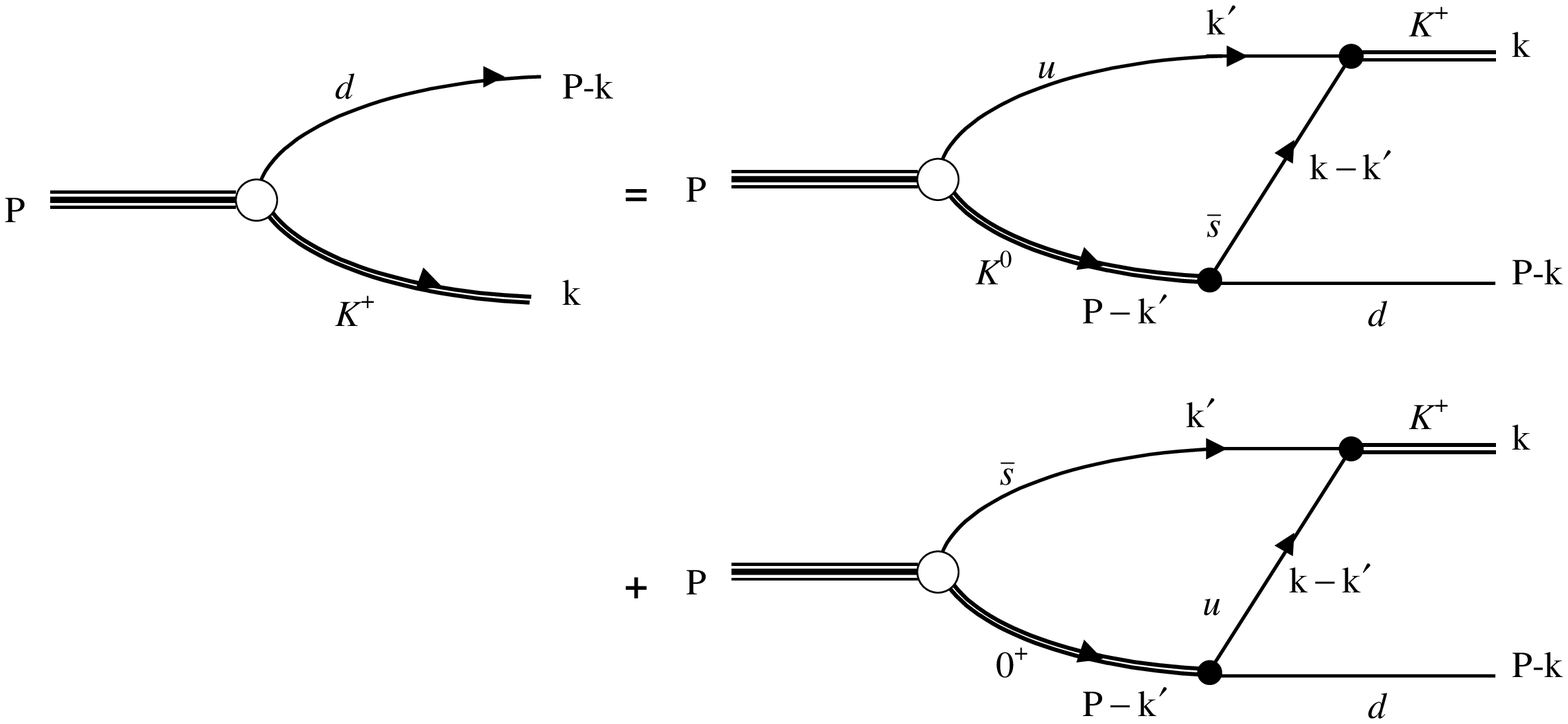}%
\caption{The equation for the triquark vertex function for the
$K^+$ and $d$ quark component.}
\end{figure}

\begin{figure}
\includegraphics[bb=20 230 340 460, angle=0, scale=0.8]{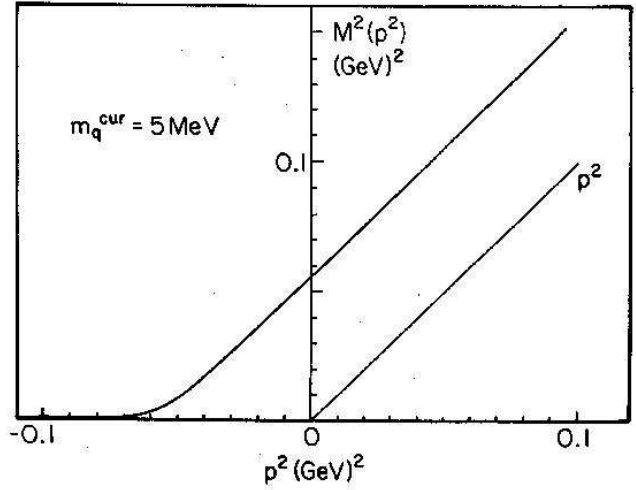}%
\caption{The square of the dynamical mass, $M^2(p^2)$, for up and
down quarks. Here we chose $m_q^{cur}=5$ MeV. [For $p^2>0,
M^2(p^2)=p^2+\kappa^2$.]}
\end{figure}

\section{Quark Propagator in The Presence of a Gluon Condensate}

In an earlier work we discussed quark propagator in the presence
of a condensate of the form $<g^2A^a_\mu A^\mu_a>$ which has
recently been shown to be the Landau gauge version of a more
general gauge invariant expression. We have discussed quark
propagation in the presence of such a condensate treating the
vacuum as a random medium of gluon fields [22]. It is found that
the quark propagator has no on-mass-shell poles indicating that
the quark cannot propagate over extended distances. As an example,
we show one of the momentum-dependent mass functions obtained in
our model in Fig. 5. It may be seen that the equation
$p^2-M^2(p^2)=0$ has no solution. In our work we have modified the
solution shown in Fig. 5 to have a constituent mass value of 350
MeV for the up (or down) quark for spacelike $k^2$ and for a small
region of timelike $k^2$ near $k^2=0$. We use a simplified form
for the momentum-dependent mass function.\be M(k^2) = [k^2 +
c^2]^{1/2} \,\,\,\,\,\,\, \mbox{for} \,\,\,\,\, k^2>m^2_q - c^2,
\ee and \be M(k^2) = m_q \,\,\,\,\,\,\, \mbox{for} \,\,\,\,\,
k^2<m^2_q - c^2, \ee with $c=0.3$ GeV and with $m_q$ being the
quark mass which we take to be 350 MeV for the up (or down) quark
and 450 MeV for the strange quark. In contrast to the result shown
in the Fig. 5, we use the constituent quark mass for $M(k^2)$ when
$k^2<m_q^2 - c^2$. That is more in keeping with standard
phenomenology, since the result shown in Fig. 5 does not capture
the behavior expected for the constituent quark mass for the
spacelike values of $k^2$.

\section{Dynamical Equations For the Triquark vertex function}

In order to construct a bound-state triquark wave function we
consider the diagrams of Figs. 2-4. In Fig. 2 we show the vertex
(open circle) for the virtual triquark decay to a $u$ quark of
momentum $P-k$ and a $K^0$ meson of momentum $k$. The $K^0$ and
$K^+$ mesons, and the diquark are taken to be on mass shell in our
formalism. (Such restrictions arise when we complete the $k'_0$
integral in the complex $k'_0$ plane.) On the right-hand side of
Fig. 2 we see the triquark component consisting of a $\bar{s}$
quark and a scalar diquark. The final state is reached by the
exchange of a $d$ quark of momentum $k-k'$. In the second figure
on the right we have a $d$ quark and a $K^+$ meson in the
intermediate state, with exchange of a $\bar{s}$ quark yielding
the final $K^0$ and $u$ quark. Similar comments pertain to the
processes shown in Figs. 3 and 4.

The triquark vertex depends upon $P$ and $k$ and has a Dirac index
$\alpha$: $\Gamma_\alpha(P,k)$. We introduce coupling constants
$g_1$ and $g_2$ which correspond to the coupling of either the
kaon or scalar diquark to their quark components. [See the
Appendix.] By completing the integral over the $k'_0$ variable, we
find we may place the kaon and the diquark on mass shell, leaving
a three-dimensional integral over $\vec{k}'$. It is also useful to
solve for $\mid\vec{k}\mid\Gamma_\alpha(P,k)$ rather than
$\Gamma_\alpha(P,k)$. We take $\vec{P}=0$ and find a bound state
at a specific value of $P^0$. The equation we solve may be written
with the Dirac indices explicit: \be
|\vec{k}|\,\Gamma_\alpha(\vec{k}) &=&
\frac{g_1g_2}{(2\pi)^2}\int^1_{-1}du\int_0^{k_{cut}}d|\vec{k'}|\,
\frac{|\vec{k}|\,|\vec{k'}|}{2E_{mes}(k')}\\\nonumber &\times &
F[(k'-k/2)^2]\,F\left[\left(k-\frac{k'+P}{2}\right)^2\right]\\\nonumber&\times&\frac{[\slr
k-\slr k'-M_1((k-k')^2)]_{\alpha\mu}}{(k-k')^2-M_1^2(k-k')^2)}
\cdot \frac{[\slr
k'+M_2(k'^2)]_{\mu\beta}}{k'^2-M_2^2(k'^2)}[\,|\vec{k'}|\,\Gamma_\beta(\vec{k'})\,].\ee
The values of $k_0$ and $k'_0$ are fixed using the on-mass-shell
conditions for the kaon and diquark. Here, $M_1$ and $M_2$ are
either $M_u, M_d$ or $M_s$ depending upon which diagram of Figs.
2-4 is being considered. The $F'$s are form factors introduced for
the diquark and kaon vertices which appear in the figures as small
filled circles.

The form factor for the final-state kaon or diquark is \be
F[(k'-k/2)^2] = \exp\left[-\frac{1}{\alpha^2}|(k'-k/2)^2|\right],
\ee with \be (k'-k/2)^2 =
(P^0-E_{mi}(\vec{k'})-E_{mf}(\vec{k})/2)^2-(\,\vec{k'}^2+\vec{k}^2/4-|\vec{k}|
|\vec{k'}| u\,),\ee where $E_{mi}(\vec{k'})$ refers to either the
kaon or diquark of momentum $\vec{k'}$ and $E_{mf}(\vec{k})$
refers to the final-state kaon or diquark.

The form factor for the intermediate-state kaon or diquark is \be
F\left[\left(k-\frac{k'+P}{2}\right)^2\right] =
\exp\left[-\frac{1}{\alpha^2}\left|\left(k-\frac{k'+P}{2}\right)^2\right|\right],
\ee with \be \left(k-\frac{k'+P}{2}\right)^2 =
(E_{mf}(\vec{k'})-P^0-E_{mi}(\vec{k})/2)^2
-(\,\vec{k}^2+\vec{k'}^2/4-|\vec{k}|\,|\vec{k'}|\,u\,).\ee

We remark that there are two terms to be considered on the
right-hand side of Eq. (3.1) when we relate that expression to the
diagrams of Figs. 2-4. There is also an implied sum over the Dirac
indices, $\mu$ and $\beta$. Since there are three decay channels
($K^0u$, $0^+\bar{s}$ and $K^+d$) and four Dirac indices
(0,1,2,3), there are twelve vertex functions to consider. If we
take $N$ points for each vertex function, we need to evaluate a
$12N$ by $12N$ matrix when searching for the bound-state
eigenvalue. In our calculation it is useful to take $\vec{k}$
along the $z$-axis so the vector $\vec{k'}$ has components
$\vec{k'}=\mid\vec{k'}\mid(\sin\theta\cos\phi,\sin\theta\sin\phi,\cos\theta)$.
We have used $u=\cos\theta$ in writing Eq. (2.1).

Once the eigenvalue is found, we may then obtain the vertex
functions or the corresponding wave function. The wave function is
\be
\psi_\alpha(\vec{k})=\frac{1}{k_0^2-\vec{k}^2-M^2(k^2)}\Gamma_\alpha(\vec{k}),
\ee where $k^0=P^0-E_{mes}(\vec{k})$ and $P^0$ is the eigenvalue.
[See Figs. 2-4.]

\section{Wave Functions of the Triquark}

In Fig. 6 we show the (unnormalized) wave functions found in our
analysis. The dashed line shows the diquark-(strange quark)
component, while the solid line exhibits the $K^0u$ and $K^+d$
components which are equal in this model. The small components of
these wave function are shown in the lower part of the figure. We
may write the four-component wave function of Eq. (3.6) as \be
\psi_s(\vec{k}) =
\left(\begin{array}{c}R_u(k)\left|s\right>\\
\vec\sigma\cdot\hat kR_l(k)\left|s\right>\end{array}\right) \,.\ee

The upper and lower components of the wave function are shown in
Fig. 6. There are three wave functions of the form of Eq. (4.1)
corresponding to the channels $0^++s$, $K^0+u$ and $K^++d$. As
noted above, the wave functions for the $K^0+u$ and $ K^++d$
components are equal in our model.

\begin{figure}
\includegraphics[bb=0 0 270 225, angle=0, scale=1]{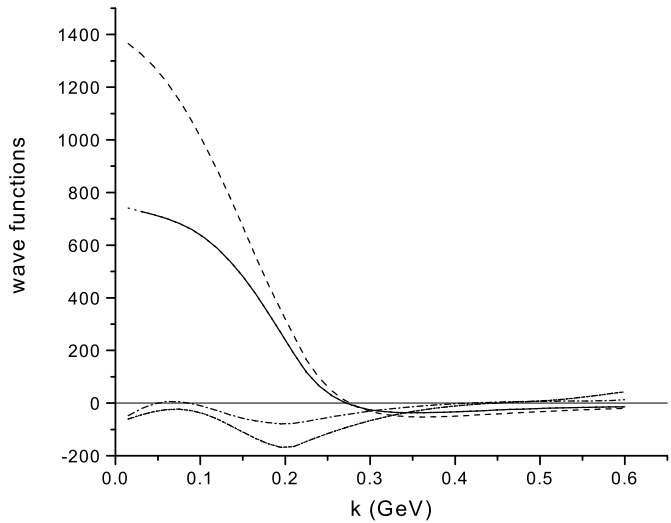}%
\caption{Wave functions of the triquark. The dashed line refers to
the diquark-quark component while the solid line represents the
(equal) components for the $K^0$ and the $u$ quark and the $K^+$
and the $d$ quark. The lower components of the wave functions are
shown in the lower part of the figure. The dash-dot curve refers
to the kaon-quark components of the triquark and the other curve
represents the lower component of the diquark-quark wave function.
In this calculation we have used $g_1=g_2=11.21, \alpha=0.2,
k_{max}=0.6$ GeV and found $P^0=0.81$ GeV. }
\end{figure}

\section{Discussion}

Our interest in triquark structure is related to the
diquark-triquark model of pentaquark structure [14-17]. As stated
earlier, it is not clear that pentaquarks exist because of various
contradictory results obtained in experimental studies. Recent
work of Karliner and Lipkin [24] appears to be quite important for
the interpretation of experimental searches for the pentaquark.
These authors claim that [24]: "Significant signal-background
interference effects can occur in experiments like $\gamma
N\rightarrow \bar{K}\Theta^+$ as a narrow $I=0$ resonance in a
definite final state against a non-resonant background, with an
experimental resolution coarser than the expected resonance width.
We show that when the signal and background have roughly the same
magnitude, destructive interference can easily combine with a
limited experimental resolution to completely destroy the
resonance signal. Whether or not this actually occurs depends
critically on the yet unknown phase of the $I=0$ and $I=1$
amplitudes ...".

In the present work we have introduced a model of triquark
structure. In order to carry out our calculation we have used a
quark self-energy that does not give rise to on-mass-shell poles.
Similar results for gluon propagation in the presence of a
condensate are presented in Ref. [25], where it is shown that the
gluon is also a nonpropagating mode in the presence of the gluon
condensate.

It would be desirable to improve the model presented in our work
and to see if there are other useful applications of the quark
propagator used in this work. Whether our triquark model may be
used in a more detailed description of the pentaquark than that we
have presented previously remains to be seen.

\appendix
\renewcommand{\theequation}{A\arabic{equation}}
  \setcounter{equation}{0}  
  \section{}

In order to calculate the normalization factor for our kaon or
diquark we consider the diagram shown in Fig. 7. We define \be
N&=& -\mbox{Tr}\int\frac{d^4k}{(2\pi)^4}\,\frac{\slr p-\slr
k+M_1[(p-k)^2]}{(p-k)^2-M_1^2[(p-k)^2]}\,\,\frac{\slr
k+M_2(k^2)}{k^2-M_2^2(k^2)}\\\nonumber&&\times(\gamma\cdot
\hat{n})\,\frac{\slr k+M_2(k^2)}{k^2-M_2^2(k^2)}\,\,f^2(p,k), \ee
where $\hat{n}=(1,0,0,0)$. (We may also identify $g=1/\sqrt{N}$ as
the effective coupling constant at the kaon or diquark vertex.) In
Eq. (A1) $f(p,k)$ is a form factor defined at the kaon or diquark
vertex. When $\vec{p}=0$, we have \be f(p,k) =
\exp\left[-\frac{1}{\alpha^2}\,|(p_0/2-k_0)^2-\vec{k}^2|\,\right].\ee

In order to calculate $N$ of Eq. (A1) we need the value of the
trace \be \mbox{Trace}&=&\mbox{Tr}{[\slr p-\slr
k+M_1[(p-k)^2][\slr k+M_2(k^2)](\gamma\cdot\hat{n}) [\slr
k+M_2(k^2)]}\\\nonumber
&=&8k_0^2p_0-4(k_0^2-\vec{k}^2)p_0+4p_0M_2^2(k^2)\\\nonumber
&&-4k_0(k_0^2-\vec{k}^2)-4k_0M_1^2(k^2)+8k_0M_1[(p-k)^2]M_2(k^2).\ee

We obtain \be N
&=&-\frac{1}{4\pi^3}\int_{-k^0_{max}}^{k^0_{max}}dk_0\int_0^{k_{max}}\vec{k}^2\,d|\vec{k}|
\left(\frac{1}{(p_0-k_0)^2-\vec{k}^2-M_1[(p-k)^2]}\right)\\\nonumber
&&\cdot f^2(p_0,k)\cdot
\left(\frac{1}{k_0^2-\vec{k}^2-M_2(k^2)}\right)\cdot(\mbox{Trace}).\ee
Here $k$ is a four-vector $k=(k_0,\vec{k})$. In our analysis we
put $k_{max}=k^0_{max}=0.6$ GeV. (For the diquark, we find
$1/\sqrt{N}=11.21$. We use the same value for the kaon since that
value is only $1.5\%$ greater than $1/\sqrt{N}$ for the diquark.)

\begin{figure}
\includegraphics[bb=130 465 480 660, angle=0, scale=1]{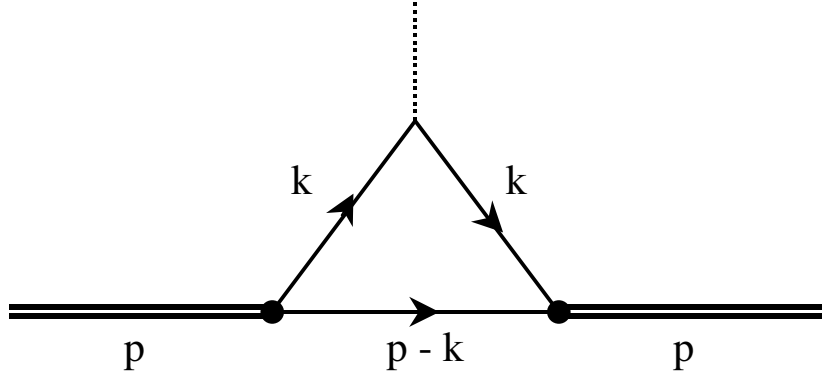}%
\caption{Diagram used in the calculation of the normalization
factor of the kaon or scalar diquark is shown. (See the
Appendix.)}
\end{figure}



\end{document}